\documentclass[12pt]{article}
\usepackage[english]{babel}
\usepackage[utf8]{inputenc}
\usepackage[T1]{fontenc}

\def\hg{\hat{g}}

\def\bAi{(\bA^{-1})}

\def\hPi{\hat{\Pi}}
\def\ot{\overline{T}}

\usepackage{amsmath}

\def\bC{\mathbf{C}}

\usepackage{amsfonts}

\usepackage{url}
\usepackage[dvips]{graphicx}
\usepackage{calc}
\usepackage{subfigure}
\usepackage{multirow}

\def\mL{\mathcal{L}}
\def\mH{\mathcal{H}}

\def\bA{\mathbf{A}}

\def\tu{\tilde{u}}

\begin{document}
	\begin{titlepage}
		\begin{center}
			{\Large{ \bf Some Remarks about  Modified  Eddington Gravity}}
			
			\vspace{1em}  
			
			\vspace{1em} J. Kluso\v{n} 			
			\footnote{Email addresses:
				klu@physics.muni.cz  }\\
			\vspace{1em}
			\textit{Department of Theoretical Physics and
				Astrophysics, Faculty of Science,\\
				Masaryk University, Kotl\'a\v{r}sk\'a 2, 611 37, Brno, Czech Republic}
			
			\vskip 0.8cm
			
			%
			%
			%
			%
			%
			%
			
			\vskip 0.8cm
			
		\end{center}

		\begin{abstract}
		In this short note we find covariant canonical formulation of modified  Eddington gravity action coupled to scalar field. We also discuss limitation of this formulation and suggest its possible generalization.
		\end{abstract}
		
		\bigskip
		
	\end{titlepage}
	
	\newpage

\section{Introduction and Summary}\label{first}
The gravitational equations of motion are usually derived from Einstein-Hilbert action when the metric of space-time $g_{ab}$ is considered as fundamental variable. In this case we obtain 
equations of motion by standard variation procedure even if it is not well defined due to the fact that Einstein-Hilbert action 
contains second derivative of metric. Of course, this problem can be solved by adding boundary term to the action that however
crucially depends on the choice of the boundary surface
\cite{Dyer:2008hb,Parattu:2015gga,Chakraborty:2016yna,Jubb:2016qzt}. Even if these boundary terms do not affect equations of motion they are important
for thermodynamic interpretations of gravity
\cite{Chakraborty:2019doh,Padmanabhan:2013nxa}.

Due to the problems with Einstein-Hilbert action mentioned above  there were attempts to find alternative formulations of theory of gravity. Such a proposal was originally suggested by Eddington
\cite{Eddington}
 who introduced gravitational Lagrangian 
$\mL=\sqrt{|\det R_{ab}|}$ where $R_{ab}(\Gamma)$ is standard Ricci 
tensor defined with the connection $\Gamma^a_{bc}$ that is fundamental variable of Eddington gravity. Then performing variation of the action we obtain equations of motion that are equivalent to General Relativity equations of motion with non-zero cosmological constant and hence Eddington gravity is well defined
alternative to Einstein-Hilbert action in case of the absence of matter. 

In fact, inclusion of matter into Eddington gravity was an unsolved
problem so far.  The most popular approach is in the extension of original Eddington form of gravity to so called Born-Infeld inspired gravity. First attempt for Born-Infeld gravity, not directly related to Eddington one,  was presented in  \cite{Deser:1998rj} where metric was fundamental degree of freedom however it was shown in the same paper that this formulation is plagued with ghost due to the fact that this action contains higher order derivatives of metric. Then it was recognized in
\cite{Vollick:2003qp,Vollick:2005gc,Banados:2010ix}
 that   Born-Infeld gravity should be formulated in the first order formalism with  connection as dynamical variable 
 which closely follows Eddington proposal.
   Variation of Born-Infeld  action with respect to metric gives an algebraic equation that allows us to express metric as function of matter degrees of freedom and connection, at least in principle, for review and extensive discussion, see \cite{BeltranJimenez:2017doy}. Then inserting the metric  back to the action and performing variation with respect to connection we obtain  gravitational equations of motion
    which however are not fully  equivalent to the equations of motions derived from Einstein-Hilbert action but they contain new important modifications.

In \cite{Chakraborty:2020yag} new interesting modification of Eddington gravity was proposed where the action contains matter contribution through specific form of stress energy tensor. The dynamical degrees of freedom are  connection and matter. It is important to stress  that this action still depends on metric through the stress energy tensor for matter however it was argued in \cite{Chakraborty:2020yag} that the metric should be considered as background field which is not varied when we search for equations of motion. In fact, even if we treat this field as non-dynamical without kinetic term the variation of this action with respect to them will  lead to inconsistent results. However then it was argued in \cite{Chakraborty:2020yag} that despite of the fact that we do not consider metric as dynamical, it should be compatible with connection.  

Since  proposed inclusion of matter into Eddington action 
\cite{Chakraborty:2020yag}  is new and interesting 
we mean that it deserves to be studied further. In particular, we would like to show that equations of motion of gravity are really derived from this modified Eddington action using covariant canonical formulation of this theory
 (also known  as   Weyl-De Donder theory
\cite{DeDonder,Weyl}).  
The key point of Weyl-De Donder theory  is that   Hamiltonian
density depends on conjugate momenta $p^\alpha_M$ which are variables conjugate to $\partial_\alpha \phi^M$. In other words we tread all partial derivatives on the equal footing which clearly preserves diffeomorphism invariance. This approach is also known as
multisymplectic field theory, see for example
\cite{Struckmeier:2008zz,Kanatchikov:1997wp,Forger:2002ak}. Covariant canonical formalism is very useful for analysis of complicated covariant systems as for example Born-Infeld gravity 
and hence also modified Eddington gravity as we will demonstrate in this paper. We perform our analysis with the case when the matter is represented by collection of scalar fields that allows us to find exact form of covariant Hamiltonian. Generally it would be very difficult to find corresponding Hamiltonian due to the complicated form of stress energy tensor for various content of matter.  In our case we find covariant Hamiltonian and also determine equations of motion. We show that they can be solved as in case of Born-Infeld inspired gravity when we introduce new symmetric tensor which is compatible with connection. Then we will
argue, following
\cite{Chakraborty:2020yag} that it is natural to identify this tensor with the metric. 

Despite of the fact that covariant canonical formalism provides
elegant formulation of  modified Eddington gravity there is still an open question of its practical application. In case, when the gravity has metric compatible connection, covariant canonical Hamitonian contains boundary term that has natural thermodynamics interpretation \cite{Kluson:2020tzn,Parattu:2013gwa}. On the other hand such a term is absent in case of Eddington gravity and hence it is not clear how thermodynamics quantities can be defined. 

As the next important issue of modified  Eddington gravity is presence of the metric in  action which however is defined as background field without its own dynamics. We mean that it would be more natural to vary this field as well but this cannot be performed in the original formulation of modified  Eddington gravity. For that reason we propose its form which is based on Born-Infeld formulation of gravity
\cite{Banados:2010ix,BeltranJimenez:2017doy,Vollick:2003qp,Vollick:2005gc} however where main difference from previous attempts how to incorporate matter is that the matter contribution is given by specific combination of stress energy tensor in the same way as 
in \cite{Chakraborty:2020yag}. Then performing variation with respect to metric tensor we find algebraic equations of motion that  can be solved for metric and that now contain additional term proportional to the variation of the stress energy tensor with respect to metric. We consider two examples that can be explicitly solved. The first one when the matter is represented by collection of scalar fields has simple solution and resulting equations of motion have the form as general relativity equations of motion with modified potential for scalar field. The second example corresponds to perfect fluid and we again show that the equations of motion for metric has explicit solution. Again we find that resulting equations of motion have formally the same form as in case of Einstein-Hilbert action where however the matter is represented by fluid with stress energy tensor that has the same form as perfect fluid with complicated space-time dependence of energy density and pressure. 

Finally in order to gain more physical insight into modified Eddington gravity we consider its formulation with the help of auxiliary metric in the similar way as in Born-Infeld gravity
\cite{BeltranJimenez:2017doy}.  Then  when we express equations of motion 
with the help of auxiliary metric we find that they have the same form as follow from Einstein-Hilbert action which ensures that
modified stress energy tensor is conserved.  

This paper can be extended in many directions. It would be interesting to study cosmological solutions when theory is expressed in terms of auxiliary metric. It would be also useful to study in more details relation between fundamental and auxiliary metric in  modification Eddington gravity. We hope to return to these problems in future.

This paper is organized as follows. In the next section (\ref{second}) we review  modified  Eddington gravity and we study its covariant canonical formulation. In section (\ref{third}) we consider  modification of this construction when the metric is dynamical variable. Finally in section (\ref{fourth}) we analyze this theory formulated with the help of auxiliary metric.

\section{Modified Eddington Gravity}\label{second}
In this section we introduced an action for modified Eddington gravity as was proposed in 
\cite{Chakraborty:2020yag}
\begin{equation}\label{Padmanaction}
	S=	\int d^4x\sqrt{\det (-(R_{(ab)}-\kappa\ot_{ab})}
	 \ , 
\end{equation}
where $x^a,a,b,c,\dots=0,1,2,3$ label  points in four dimensional space-time and  where
\begin{eqnarray}
&& 
	R_{ab}=\partial_m \Gamma^m_{ab}-\partial_a \Gamma^m_{bm}
	+\Gamma^m_{mn}\Gamma^n_{ab}-\Gamma_{an}^m\Gamma^n_{mb} 
	\ , \nonumber \\	
&&	R_{(ab)}=\frac{1}{2}(R_{ab}+R_{ba}) \ , 
\nonumber \\
&&		\ot_{ab}=T_{ab}-\frac{1}{2}T g_{ab} \ , \quad \kappa=\frac{1}{M_p^2}=8\pi G_N \ , \nonumber \\
\end{eqnarray}
where $T_{ab}$ is stress energy tensor for matter that explicitly depends on metric $g_{ab}$ and where $T\equiv g^{ab}T_{ab}$.
We further presume that the connection is symmetric 
\begin{equation}
	\Gamma^c_{ab}=\Gamma^c_{ba} \ 
\end{equation}
even if the present construction can be generalized to non-symmetric connection as well \cite{Chakraborty:2020yag}.
Note that $R_{ab}$ is not symmetric in $a,b$ indices due to the presence of the term $\partial_a \Gamma^m_{bm}$ which is not symmetric unless $\Gamma^a_{bc}$ is Christoffel connection.

 The crucial point of the action (\ref{Padmanaction}) is that it  depends on metric $g_{ab}$ through the stress energy tensor $\ot_{ab}$. In \cite{Chakraborty:2020yag} the metric was treated as variable which is not varied when we determine corresponding equations of motion. 
 In other words it appears as spectator in the theory however its presence is necessary due to the fact that it is not possible to formulate stress energy tensor for matter that does not depend on metric.
 
 Before we proceed to the  covariant canonical formulation of the action (\ref{Padmanaction})  we should be more specific about  matter part of the action. Generally stress energy tensor is more complicated than corresponding action we start with. This fact  implies that covariant canonical formulation will be rather involved. An exceptional case corresponds to the situation when we have  
 a collection o $N$ scalar fields $\phi^A, A=1,\dots,N$ with the action 
\begin{equation}\label{actSmatt}
	S_{matt}=-\frac{1}{2}\int d^4x\sqrt{-g}(g^{ab}\partial_a\phi^A\partial_b\phi^B K_{AB}-V(\phi)) \ . 
\end{equation}
It is important to stress that this action should be considered
as useful tool for definition of the stress energy tensor rather than fundamental action for scalar field. In fact, using the canonical definition of the stress energy tensor
\begin{equation}
T_{ab}=-\frac{2}{\sqrt{-g}}\frac{\delta S_{matt}}{\delta g^{ab}}	
\end{equation}
we obtain from (\ref{actSmatt}) following stress energy 
tensor for $N$ scalar fields
\begin{equation}
	T_{ab}=
	\partial_a \phi^A\partial_b \phi^B K_{AB}	-\frac{1}{2}g_{ab}(g^{cd}\partial_c \phi^A\partial_d\phi^BK_{AB}-V) \ .  
\end{equation}
Since $T=g^{ab}T_{ab}=2V-g^{ab}\partial_a \phi^A\partial_b \phi^B $
we obtain that $\ot_{ab}$ is equal to
\begin{eqnarray}\label{Tscal}
	\ot_{ab}=T_{ab}-\frac{1}{2}Tg_{ab}=
	\partial_a \phi^A\partial_b \phi^BK_{AB}-\frac{1}{2}V g_{ab} \ .
	\nonumber \\	
\end{eqnarray}
The action  (\ref{Padmanaction}) with stress energy tensor 
given in (\ref{Tscal}) will be starting point for 
 covariant canonical formulation of modified Eddington gravity.

\section{Covariant 
	Canonical Formalism of Modified Eddington Gravity}
\label{third}
In this section we proceed to the covariant canonical formalism of the action (\ref{Padmanaction}) coupled to collection of $N$ scalar fields. Recall that in standard non-covariant canonical formalism canonically conjugate momentum is defined as 
 partial derivative of Lagrangian density with respect to the time derivative of corresponding field. In case of covariant canonical momentum we define it as  derivative of Lagrangian density with respect to all partial derivatives of corresponding field. For that reason all canonical conjugate momenta contain upper space-time index. Explicitly, in case of the action (\ref{Padmanaction}) we obtain
\begin{eqnarray}
&&	\Pi^{abd}_{\quad c}=\frac{\partial \mL}{\partial(\partial_d \Gamma^c_{ab})}=
\frac{1}{2}\sqrt{\det\bA}(\bA^{ab}\delta_c^d-\frac{1}{2}
	(\bA^{da}\delta_c^b+\bA^{db}\delta_c^a))  \ , 
	\nonumber \\
&&p_A^c=\frac{\partial \mL}{\delta \partial_c \phi^A}=-\kappa
\sqrt{-\det \bA}(\bA^{-1})^{cd}\partial_d\phi^BK_{BA} \ , \nonumber \\
\end{eqnarray}
where $\bA_{ab}=R_{(ab)}-\kappa \ot_{ab}$ and $\bAi^{ab}$ is inverse matrix $\bA_{ab}\bAi^{bc}=\delta_a^c$. Now with the help of corresponding conjugate momenta we define    Hamiltonian density  as
\begin{eqnarray}
	\mH=\Pi^{abd}_{\quad c}
	\partial_d \Gamma^c_{ab}+p^c_A\partial_c \phi^A-\mL \ .
	\nonumber \\
\end{eqnarray}
To proceed further we use the fact that 
\begin{eqnarray}
	\Pi^{abd}_{ \quad c}(\Gamma^c_{am}\Gamma^m_{bd}+
	\Gamma^c_{bm}\Gamma^m_{ad})=
	\frac{1}{2}
	\sqrt{-\det\bA}
	\bA^{ab}[\Gamma^d_{am}\Gamma^m_{bd}-
	\Gamma_{ab}^m\Gamma_{mn}^n]
	\nonumber \\	
\end{eqnarray}
and also 
\begin{eqnarray}
&&	\Pi^{abd}_{\quad c}\Gamma^m_{ab}\Gamma_{md}^c=
	\frac{1}{2}
	\sqrt{-\det\bA}\bA^{ab}[\Gamma^m_{ab}\Gamma_{mc}^c-
	\Gamma^m_{an}\Gamma^n_{mb}] \ , 
	\nonumber \\
&&	\Pi^{abd}_{\quad c}\partial_d \Gamma^c_{ab}=
	\frac{1}{2}\sqrt{-\det\bA}\bA^{ab}[\frac{1}{2}(\partial_d \Gamma^d_{ab}+\partial_d^d
	\Gamma_{ba})-
	\frac{1}{2}(\partial_b\Gamma^m_{ma}+\partial_a\Gamma^m_{mb})] \ , 
	\nonumber \\	
&&p_A^c\partial_c\phi^A
=-\kappa\sqrt{-\det\bA}(\bA^{-1})^{ab}(\ot_{ba}
+\frac{1}{2}g_{ba}V) \ . 
\nonumber \\
\end{eqnarray}
Then we can write 
\begin{eqnarray}
&&	\Pi^{abd}_{\quad c}\partial_d \Gamma^c_{ab}+p_A^c\partial_c\phi^A
	=2\sqrt{-\det\bA}+\frac{1}{2}	\Pi^{abd}_{ \quad c}(\Gamma^c_{am}\Gamma^m_{bd}+\nonumber \\
	&&+
	\Gamma^c_{bm}\Gamma^m_{ad}
	-\Gamma^m_{ab}\Gamma^c_{md})
	-\frac{1}{2}\kappa\sqrt{-\det\bA}\bAi^{ab}(\ot_{ab}+g_{ab}V) \  \nonumber \\
\end{eqnarray}
and hence Hamiltonian density is equal to
\begin{eqnarray}
&&	\mH=
	\Pi^{abd}_{\quad c}\partial_d \Gamma^c_{ab}+p_A^c\partial_c\phi^A-\mL=
\nonumber \\
&&	=\sqrt{-\det\bA}+\frac{1}{2}	\Pi^{abd}_{ \quad c}(\Gamma^c_{am}\Gamma^m_{bd}+
\Gamma^c_{bm}\Gamma^m_{ad}
-\Gamma^m_{ab}\Gamma^c_{md})-\nonumber \\
&&-\frac{1}{2}\kappa\sqrt{-\det\bA}\bAi^{ab}(K_{ab}+\frac{1}{2}g_{ab}V)  \ , K_{ab}=\partial_a\phi^A
K_{AB}\partial_b \phi^B \ . 
\nonumber \\
\end{eqnarray}
However this is not final form of the Hamiltonian density since
it is not function of canonical variables. 
In fact, following 
\cite{Kluson:2024kbc}
 we use the fact that 
\begin{equation}
	\Pi^{abc}_{\quad c}=\frac{3}{2}
	\sqrt{-\det\bA}\bAi^{ab}
\end{equation}
so that
\begin{equation}
	\det \Pi^{abc}_{\quad c}=
	\left(\frac{3}{2}\right)^4\det\bA 	\ . 
\end{equation}
Let us further introduce matrix $\hPi_{ab}$ as matrix inverse to $\Pi^{abc}_{\quad c}$
\begin{equation}
	\Pi^{abc}_{\quad c}\hPi_{bd}=\delta^a_d \ . 
\end{equation}
Then we obtain that the 
 Hamiltonian is equal to
\begin{eqnarray}\label{mHfinal}
&&	\mH=\frac{4}{9}\sqrt{-\det \Pi^{abc}_{\quad c}}
	+\frac{1}{2}	\Pi^{abd}_{ \quad c}(\Gamma^c_{am}\Gamma^m_{bd}+
	\Gamma^c_{bm}\Gamma^m_{ad}
	-\Gamma^m_{ab}\Gamma^c_{md}) -\nonumber \\
&&-\frac{\kappa}{6}\Pi^{abc}_{\quad c}g_{ab}V-\frac{3}{4\kappa}p^a_A\hPi_{ab}p^b_BK^{AB} \ .
\nonumber \\
\end{eqnarray}
Note that the boundary term is absent in this Hamiltonian density
which is a consequence of the fact that fundamental degrees of freedom is connection and hence an action function of first order derivatives of connection only. 

With the help of this Hamiltonian density (\ref{mHfinal})   we obtain canonical form of the action 
\begin{equation}
	S_{can}=\int d^4x (p_A^a\partial_a\phi^A+\Pi^{abd}_{\quad  c}\partial_d\Gamma_{ab}^c-\mH)
\end{equation}
and canonical equations of motion are derived by variation of this action 
\begin{eqnarray}
&&	\delta S_{can}=\int d^4x (\delta p^a_A\partial_a\phi^A-\partial_a p^a_A\delta\phi^A+
	\delta\Pi^{abd}_{\quad c}\partial_d\Gamma^c_{ab}-\partial_d\Pi^{abd}_{\quad c}
	\delta\Gamma^c_{ab}-\nonumber \\
&&	-\frac{\delta \mH}{\delta p^a_A}\delta p^a_A-
	\frac{\delta \mH}{\delta \phi^A}\delta \phi^A-
	\frac{\delta \mH}{\delta \Pi^{abd}_{\quad c}}\delta\Pi^{abd}_{\quad c}-\frac{\delta \mH}{
		\delta \Gamma^c_{ab}}\delta \Gamma^c_{ab})
	\nonumber \\
\end{eqnarray}
so that we obtain following set of equations of motion 
\begin{eqnarray}\label{eqmon}
&&		\partial_a\phi^A+\frac{3}{2\kappa}\hPi_{ab}p^b_BK^{AB}=0 \ , 
	\nonumber \\
&&-\partial_a p^a_A+\frac{\kappa}{6}\Pi^{abc}_{ \quad c}g_{ab}\frac{\delta V}{\delta \phi^A}=0 \ , 
\nonumber \\
&&\partial_d\Gamma^c_{ab}
-\frac{2}{9}\sqrt{-\det \Pi^{abc}_{\quad c}}\hPi_{ab}\delta_d^c
-\nonumber \\
&&-\frac{1}{2}(\Gamma^c_{am}\Gamma^m_{bd}+
\Gamma^c_{bm}\Gamma^m_{ad}
-\Gamma^m_{ab}\Gamma^c_{md})+\frac{\kappa}{6}\delta^c_dg_{ab}V-\frac{3}{4\kappa}
p^m_A\hPi_{ma}\hPi_{nb}p^n_BK^{AB}=0 \ , \nonumber \\
\nonumber \\
&&-\partial_d\Pi^{abd}_{\quad c}-\frac{1}{2}
(\Pi^{amn}_{\quad c}\Gamma^b_{mn}+\Pi^{bmn}_{\quad c}\Gamma^a_{mn}+
\Pi^{amb}_{\quad b}\Gamma^n_{mc}+
\Pi^{bma}_{\quad n}\Gamma^n_{mc}-\nonumber \\
&&-\Pi^{abm}_{\quad n}\Gamma^n_{mc}
-\frac{1}{2}\Pi^{mnb}_{\quad c}\Gamma^a_{mn}-\frac{1}{2}
\Pi^{mna}_{\quad c}\Gamma^b_{mn})=0
\nonumber \\
\end{eqnarray}
using the fact that 
\begin{eqnarray}
		\frac{\delta \hPi_{mn}}{\delta \Pi^{abd}_{\quad c}}=-\frac{1}{2}
		(\hPi_{ma}\hPi_{bn}+\hPi_{mb}\hPi_{bm})\delta^d_c
\ . 		\nonumber 
		\\	
\end{eqnarray}
Let us analyze these equations  in more details. The first two determine dynamics of scalar fields $\phi^A$. The last two describes evolution of connection and conjugate momenta where the metric is still non-dynamical field and there is no relation between  connection and
metric. On the other hand the equation (\ref{eqmon}) should be equivalent to equations of motion derived from Einstein-Hilbert action. To see this we will follow analysis presented in 
\cite{Kluson:2024kbc} and deduce 
that  $\Pi^{abd}_{\quad c}$ can be written as 
\begin{equation}\label{Piabc}
\Pi^{abd}_{\quad d}=\sqrt{-\det\bC}
(K \bC^{ab}\delta^d_c+L(\bC^{ad}\delta_c^b+
\bC^{bd}\delta_c^a)) \ , 
\end{equation}
where we introduced new independent tensor $\bC^{ab}$ with its inverse $\bC_{ab}$. Further, $K,L$ are unknown constants that will be determined below. 
From (\ref{Piabc}) we obtain
\begin{eqnarray}
	\partial_d\Pi^{abd}_{\quad c}=
	\sqrt{-\det \bC}(K\partial_d \bC^{ab}\delta^d_c+
	\partial_d \bC^{ad}\delta^b_c+
	\partial_d\bC^{bd}\delta^a_c)+\nonumber \\
	+\frac{1}{2}\sqrt{-\det \bC}\partial_d
\bC_{mn}\bC^{mn}(K\bC^{ab}\delta^d_c+L(\bC^{ad}\delta^b_c+\bC^{bd}\delta^a_c))	
\end{eqnarray}
and hence the third equation in (\ref{eqmon}) is equal to
\begin{eqnarray}\label{eqPiabc}
&&\sqrt{-\bC}K(\partial_d \bC^{ab}+\bC^{am}\Gamma_{md}^b+
	\Gamma^a_{md}\bC^{mb})+\nonumber \\
&&	+\sqrt{-\det \bC}\bC^{ab}(\frac{1}{2}K\partial_d \bC_{mn}\bC^{mn}+
	(L-\frac{K}{2})\Gamma^m_{md})+\nonumber \\
&&	+\sqrt{-\bC}(L\partial_c \bC^{ac}-L\frac{1}{2}\partial_c \bC^{mn}\bC_{mn}\bC^{ac}+\frac{1}{2}(L-\frac{1}{2}K)\bC^{mn}
	\Gamma_{mn}^a)\delta_d^b+\nonumber \\
&&+	\sqrt{-\bC}(L\partial_c \bC^{bc}-L\frac{1}{2}\partial_c \bC^{mn}\bC_{mn}\bC^{bc}+\frac{1}{2}(L-\frac{1}{2}K)\bC^{mn}
	\Gamma_{mn}^a)\delta_d^a=0\ . \nonumber \\
\end{eqnarray}
The equation (\ref{eqPiabc}) can be solved when we impose 
compatibility condition between the matric $\bC_{ab}$ and connection 
$\Gamma^c_{ab}$  
\begin{equation}\label{comcon}
	\partial_d \bC^{ab}+\bC^{am}\Gamma^b_{md}+\Gamma^a_{md}\bC^{mb}=0 \ 
\end{equation}
that firstly implies that an expression on the first line in (\ref{eqPiabc}) vanishes. Then if we multiply (\ref{comcon}) with $\bC_{ab}$ and sum over $a$ and $b$ we get
\begin{equation}
	\partial_d \bC^{mn}\bC_{mn}=-2\Gamma^m_{md} 
\end{equation}
that also implies
\begin{equation}
	\partial_d \bC_{mn}\bC^{mn}=2\Gamma^m_{md} 
\end{equation}
as follows from the fact that 
 $\partial_d \bC_{mn}\bC^{nm}+\bC_{nm}\partial_d\bC^{nm}=0$. If we use  this formula on the second line in (\ref{eqPiabc})  and demand that it vanishes separately we obtain 
 the result 
\begin{eqnarray}\label{conKL}
	 L=-\frac{K}{2} \ . 
	\nonumber \\
\end{eqnarray}
Finally expressions on the third and fourth lines in (\ref{eqPiabc}) identically vanish 
using (\ref{comcon}) and (\ref{conKL}).
 In other words we showed that equations of motion for $\Pi^{abc}_{\quad d}$ are solved by introducing new geometrical object  $\bC_{ab}$  that is compatible 
with connection $\Gamma^a_{bc}$ (\ref{comcon}).

Finally we should consider equations of  motion for $\Pi^{abd}_{\quad c}$ that we rewrite into the form 
\begin{eqnarray}\label{eqabdPi}
&&	\partial_d\Gamma^c_{ab}
	-\frac{2}{9}\sqrt{-\det \Pi^{abc}_{\quad c}}\hPi_{ab}\delta_d^c
	-\nonumber \\
&&	-\frac{1}{2}(\Gamma^c_{am}\Gamma^m_{bd}+
	\Gamma^c_{bm}\Gamma^m_{ad}
	-\Gamma^m_{ab}\Gamma^c_{md})+\frac{\kappa}{6}\delta^c_dg_{ab}V-\frac{3}{4\kappa}
	p^m_A\hPi_{ma}\hPi_{nb}p^n_BK^{AB}=0  \nonumber \\
	\end{eqnarray}
Performing contraction over $d$ and $c$ in (\ref{eqabdPi}) we obtain
\begin{eqnarray}
&&	\partial_c\Gamma^c_{ab}
	-\frac{8}{9}\sqrt{-\det \Pi^{abc}_{\quad c}}\hPi_{ab}
	-\nonumber \\
&&	-\frac{1}{2}(\Gamma^c_{am}\Gamma^m_{bc}+
	\Gamma^c_{bm}\Gamma^m_{ac}
	-\Gamma^m_{ab}\Gamma^c_{mc})+\frac{4\kappa}{6}g_{ab}V-\frac{12}{4\kappa}
	p^m_A\hPi_{ma}\hPi_{nb}p^n_BK^{AB}=0\ . \nonumber \\
\end{eqnarray}
If we combine this equation with  (\ref{eqabdPi})  we obtain 
\begin{eqnarray}\label{eqhelp}
&&	\partial_c\Gamma^c_{ab}-\partial_a\Gamma^m_{bm}
+\Gamma^m_{ab}\Gamma^n_{mn}-\Gamma^n_{am}\Gamma^m_{nb}-
\frac{2}{3}\sqrt{-\det\Pi^{abc}_{\quad c}}
\hPi_{ab}+\nonumber \\
&&+\kappa g_{ab}V-\frac{9}{4\kappa}p^m_A \hPi_{ma}\hPi_{nb}p^n_B
K^{AB}=0 \ . \nonumber \\
\end{eqnarray}
Now since 
\begin{equation}
	\Pi^{abd}_{\quad d}=3K\bC^{ab}\sqrt{-\det \bC} \ , \quad 
	\hPi_{ab}=
	\frac{1}{3K\sqrt{-\det\bC}}\bC_{ab}
\end{equation}
we finally get that (\ref{eqhelp})  has the form
\begin{eqnarray}
	R_{ab}-2K\bC_{ab}+\frac{\kappa}{2} g_{ab}V-\kappa\partial_a \phi\partial_b\phi	
	=0 \nonumber \\
\end{eqnarray}
At this point we see a striking property of modified Eddington gravity which is an existence of two metric tensors $\bC_{ab}$ and $g_{ab}$ which are appriory independent. On the other hand it was argued in  
\cite{Chakraborty:2020yag} that  $\bC_{ab}$ should be proportional to $g_{ab}$. These arguments follow  from Einstein equivalence principle which says that 
locally we can eliminate effect of gravity by transformation into inertial frame by suitable coordinate transformations. Clearly it is not possible to eliminate both $\bC_{ab}$ and $g_{ab}$ by four coordinate transformations and hence it is natural to demand that $\bC_{ab}$ is proportional to $g_{ab}$.  Alternatively, we see that at the point where potential vanishes 
the equations of motion do not depend on the metric $g_{ab}$ at all so that it is natural to say that $\bC_{ab}$ corresponds to the space-time metric and our result naturally reproduces spirit of Eddington gravity.
 

On the other hand we feel that it is not satisfactory that the action of modified Eddington gravity  depends on metric which is background field without its own dynamics. For that reason we try to improve original form of modified Eddington gravity to make metric tensor dynamical in a sense that we perform variation of  action with respect to metric too. We propose such a form of action in the next section. 

\section{Metric in Modified Eddington Gravity as Dynamical Field  }
\label{fourth}
In the original formulation of modified Eddington gravity the metric $g_{ab}$ was fixed  without corresponding variation. 
In fact, if we consider original modified Eddington  model and treat $g_{ab}$ as dynamical variable and perform variation with respect to $g_{ab}$ we will  get the equation
\begin{equation}
	\frac{\delta \ot_{ab}}{\delta g_{mn}}=0 
\end{equation}
which certainly is very strong condition. For that reason we should consider more general form of modified Eddington gravity and it turns out that the natural generalization corresponds to 
Born-Infeld inspired action 
\begin{eqnarray}\label{Padmod}
&&S=M_p^2M_{BI}^2\int d^4x \left(\sqrt{-\det \bA_{ab}}-\lambda\sqrt{-\det g}\right) \ ,
\nonumber \\
&&\bA_{ab}=g_{ab}+\frac{1}{M_{BI}^2}(R_{(ab)}-\kappa \ot_{ab}) \ . 
\end{eqnarray}
where from dimensinal reason we introduced scale  $M_{BI}$
that could be equal to $M_p$ but we keep it arbitrary. Note that (\ref{Padmod}) has the form of Born-Infeld modified action where
however matter contribution is given by $\ot_{ab}$ included into the determinant of matrix that defines Born-Infeld gravity
 \cite{BeltranJimenez:2017doy}. It is also important to introduce term $\sqrt{-g}$ into action multiplied 
by non-zero parameter   $\lambda$ since it  is crucial for consistency of theory. We should also stress that  Born-Infeld gravity with the matter part included into determinant was previously introduced in 
\cite{Vollick:2005gc} but our approach differs from the form of matter contribution.

Since $g_{ab}$ is now dynamical we obtain from (\ref{Padmod}) corresponding equations of motion 
\begin{eqnarray}\label{algeqg}
	\sqrt{-\det \bA_{ab}}\bAi^{mn}(\frac{1}{2}(\delta_m^a\delta_n^b+\delta_m^b
	\delta_n^a)-\frac{\kappa}{M_{BI}^2}\frac{\delta \ot_{mn}}{\delta g_{ab}})
	-\lambda \sqrt{-g}g^{ab}=0 \nonumber \\
\end{eqnarray}
while variation with respect to $\Gamma^c_{ab}$ gives
\begin{eqnarray}\label{eqmodGamma}
&&\nabla_c(
\sqrt{-\det \bA}\bAi^{ab})+\frac{1}{2}\nabla_m(\sqrt{-\det\bA}\bAi^{ma}\delta_c^b)+\nonumber \\
&&+\frac{1}{2}\nabla_m (\sqrt{-\det\bA}\bAi^{ mb}\delta_c^a)=0 \ . 
\nonumber \\
\end{eqnarray}
Performing contraction between $c$ and $a$ in (\ref{eqmodGamma}) we obtain  
\begin{equation}
\nabla_m(\sqrt{-\det \bA}\bA^{mb})=0
\end{equation}
that inserting back to (\ref{eqmodGamma}) gives
\begin{equation}
	\nabla_c (\sqrt{-\det \bA}\bAi^{ab})=0 \ .
\end{equation}
Since $\bA_{ab}$ is non-singular matrix the previous equation is equivalent to the condition 
\begin{equation}
	\nabla_c \bAi^{ab}=0 \  .
\end{equation}
In other words we can express $\Gamma^c_{ab}$ as
\begin{equation}
	\Gamma^c_{ab}=\frac{1}{2}
	\bAi^{cd}(\partial_a \bA_{db}+\partial_b\bA_{da}-\partial_d 
	\bA_{ab}) \ , 
\end{equation}
where however $\bA_{ab}$ should  be determined from 
(\ref{algeqg}).  We also see that it is crucial that $\lambda$ is non-zero which ensures that $\bA_{ab}$ is non-singular matrix on-shell.  Note that in the absence of matter or when stress energy tensor does not depend on $g_{ab}$ the equation (\ref{algeqg}) shows that $\bA_{ab}$ is proportional to $g_{ab}$ and we recovery standard Eddington gravity.  On the other hand 
in case when $\frac{\delta \ot_{ab}}{\delta g_{mn}}\neq 0$ the situation is more involved. Let us now study two examples where the equation 
(\ref{algeqg}) can be explicitly solved. 

The first example corresponds to  the collection of $N$ scalar fields which was studied in previous section where the tensor
 $\ot_{ab}$ is equal to 
\begin{equation}
	\ot_{ab}=\partial_a\phi^A\partial_b\phi^BK_{AB}-\frac{1}{2}V(\phi)g_{ab} \ 
\end{equation}
and hence
\begin{equation}
	\frac{\delta \ot_{mn}}{\delta g_{ab}}=-\frac{1}{4}V(\delta_m^a\delta_n^b+\delta_m^b
	\delta_n^a) 
\end{equation}
so that (\ref{algeqg}) has the form 
\begin{eqnarray}
\sqrt{-\det\bA}\bAi^{ab}(1+\frac{\kappa}{2M_{BI}^2}V)-\lambda \sqrt{-g}g^{ab}=0 \ . 
\nonumber \\
\end{eqnarray}
If we calculate determinant of this expression we obtain 
\begin{equation}
	\det\bA=\frac{\lambda^4}{(1+\frac{\kappa}{2M_{BI}^2}V)^4}\det g
\end{equation}
and hence we find that 
\begin{equation}
	\bAi^{ab}=\frac{1}{\lambda}(1+\frac{\kappa}{2M_{BI}^2}V)g^{ab} \ , \quad 
	\bA_{ab}=\frac{\lambda}{1+\frac{\kappa}{2M_{BI}^2}V}g_{ab} \ . 
\end{equation}
This is very important result that says that the matrix $\bA_{ab}$ is related to 
$g_{ab}$ by Weyl rescaling with the factor
\begin{equation}
	\Omega=\frac{\lambda}{1+\frac{\kappa}{2M_{BI}^2}V} \ .  
\end{equation}
On the other hand since $\bA_{ab}=g_{ab}+\frac{1}{M_{BI}^2}
(R_{(ab)}-\ot_{ab})$ we find equation of motion in the form 
\begin{equation}\label{eqphi}
	g_{ab}+\frac{1}{M_{BI}^2}
	(R_{ab}(\bA)-\kappa\ot_{ab})=\frac{\lambda}{1+\frac{\kappa}{2M_{BI}^2}V}g_{ab} \ , 
\end{equation}
where $R_{(ab)}=R_{ab}$ due to the fact that $\Gamma$ is Christoffel
connection for matrix $\bA_{ab}$. Note that the equation 
(\ref{eqphi}) can be rewritten into the form 
\begin{equation}
	R_{ab}(\Omega g)-\kappa\ot_{ab}=g_{ab}M_{BI}^2
(\frac{\lambda}{1+\frac{\kappa}{2M_{BI}^2}V}-1)	\ . 
\end{equation}
At this formulation the fundamental gravitational degrees of freedom correspond to metric $g_{ab}$. In fact, we can rewrite
$R_{ab}(\Omega g)$ as function of $R_{ab}(g)$ and conformal 
factor $\Omega$ and their partial derivatives using generalized
Weyl rescaling of metric. 
However there is another interesting possibility when we express
$g_{ab}$ with the help of the matrix $\bA_{ab}$
\begin{equation}
	g_{ab}=\bA_{ab}\Omega^{-1} \ 
\end{equation}
so that the equations of motion have the form 
\begin{eqnarray}
	R_{ab}(\bA)-\kappa \partial_a\phi^A\partial_b\phi^BK_{AB}
+M_{BI}^2(\frac{1}{\lambda}(1+\frac{\kappa}{2M_{BI}^2}V)^2-1)\bA_{ab}=0
\nonumber \\	
\end{eqnarray}
that can be written as 
\begin{eqnarray}\label{modeq}
&&	R_{ab}(\bA)-\frac{1}{2}R(\bA)\bA_{ab}-\kappa \tilde{T}_{ab}=0 \ , \nonumber \\
&&\tilde{T}_{ab}=	\partial_a\phi^A\partial_b\phi^BK_{AB}
+\frac{M^2_{BI}}{\kappa}\left(\frac{1}{\lambda}(1+\frac{\kappa}{2M^2_{BI}}V)^2-1\right)\bA_{ab}-\nonumber \\
&&-\frac{1}{2}\bA_{ab}
\bA^{mn}\partial_m\phi^A\partial_n\phi^BK_{AB} \ . 
\nonumber \\
\end{eqnarray}
It is important to stress that the  equations on the first line in  (\ref{modeq}) 
are standard general relativity equations of motion when the fundamental metric is $\bA_{ab}$. Note also using the fact that Einstein tensor $R_{ab}(\bA)-\frac{1}{2}R\bA_{ab}$ obeys an identity
\begin{equation}
	\nabla_a (R^{ab}-\frac{1}{2}\bAi^{ab}R)=0 \ 
\end{equation}
we immediately obtain that the stress energy tensor is conserved
\begin{equation}
	\nabla_a \tilde{T}^{ab}=0 \ . 
\end{equation}
This fact also implies that the dynamics of the scalar field is governed
by the metric $\bA_{ab}$ that is related to the metric $g_{ab}$ by Weyl rescaling.

As the second example we  consider stress energy tensor for perfect fluid 
\begin{equation}
	T^{ab}=(\rho+p)u^au^b+pg^{ab} \ , u^au^bg_{ab}=-1 \ ,
\end{equation}
where $\rho$ and $p$ are energy density and pressure of the fluid 
and $u^a$ is corresponding four-velocity. 
The trace of the stress energy tensor is equal to
\begin{equation}
	T=g_{ab}T^{ab}=-\rho+3p
\end{equation}
so that
\begin{eqnarray}
&&	\ot_{ab}
=(\rho+p)g_{ac}u^cu^d g_{db}+\frac{1}{2}(\rho-p)g_{ab} \ 
\end{eqnarray}
and hence 
\begin{equation}
\frac{\delta \ot_{mn}}{\delta g_{ab}}=\frac{1}{4}(\rho-p)(\delta_m^a\delta_n^b+\delta_m^b\delta_n^a)+
\frac{1}{2}(\rho+p)((\delta_m^a u^b+\delta_m^b u^a)u_n+u_m(\delta_n^au^b+
\delta_n^bu^a))
\end{equation}
so that the equations of motions (\ref{algeqg}) have the form 
\begin{eqnarray}\label{algeqgfluid}
&&\sqrt{-\det \bA}\bA^{ab}-\nonumber \\
&&-\frac{\kappa}{M_{BI}^2}
\sqrt{-\det \bA}
(\frac{1}{2}(\rho-p)\bAi^{ab}+(\rho+p)(u^cg_{cm}\bAi^{ma}u^b+u^au^cg_{cm}\bAi^{mb}))=\nonumber \\
&&=\lambda\sqrt{-g}g^{ab}\ . 	\nonumber \\
\end{eqnarray}
Let us presume that solution of this equation is given by following ansatz
\begin{equation}\label{bAi}
	\bAi^{ab}=Xg^{ab}+Y u^au^b \ , 
\end{equation}
where $X$ and $Y$ will be determined below. Inserting (\ref{bAi})
into the left side of (\ref{algeqgfluid}) we obtain that it is equal to
\begin{eqnarray}\label{algeqgfluid1}
&&	\frac{\sqrt{-\det g}}{X^2
	\sqrt{1-\frac{2\kappa}{M^2_{BI}}\frac{(\rho+p)}{1+\frac{\kappa}{2M^2_{BI}}
			(3\rho+5p)}}}\times \nonumber \\
&&		\times 
		\left(
Xg^{ab}+Y u^au^b-
\frac{\kappa}{M_{BI}^2}
(\frac{1}{2}(\rho-p)(Xg^{ab}+Yu^au^b)+2
(\rho+p)(X-Y)u^au^b)\right) \ ,  \nonumber \\
\end{eqnarray}
where we used  the fact that 
\begin{eqnarray}
	\sqrt{-\det \bA}=\frac{1}{\sqrt{-\det \bAi}}=
	\frac{\sqrt{-\det g}}{X^2
		\sqrt{1-\frac{2\kappa}{M^2_{BI}}\frac{(\rho+p)}{1+\frac{\kappa}{2M^2_{BI}}
				(3\rho+5p)}}}	\ . 	\nonumber \\
\end{eqnarray}
Since  the right side of the equation (\ref{algeqgfluid}) 
does not depend on $u^a$ it is natural to demand that  
terms  proportional to $u^au^b$ in (\ref{algeqgfluid1})
 vanish so that we can express $Y$ as 
\begin{eqnarray}\label{Y}
	Y=\frac{2\kappa}{M_{BI}^2}(\rho+p)X
	\frac{1}{1+\frac{\kappa}{2M_{BI}^2}(3\rho+5p)} \ . \nonumber \\
\end{eqnarray}
Then finally (\ref{algeqgfluid1}) gives 
\begin{eqnarray}\label{X}
X=\frac{(1-\frac{\kappa}{2M_{BI}^2}(\rho-p))}{\sqrt{1-\frac{2\kappa}{M^2_{BI}}\frac{(\rho+p)}{1+\frac{\kappa}{2M^2_{BI}}
			(3\rho+5p)}}}=
		\sqrt{(1-\frac{\kappa}{2M_{BI}^2}(\rho-p))(
			1+\frac{\kappa}{2M_{BI}^2}(3\rho+5p))} \ , \nonumber \\
\end{eqnarray}
so that we find that $\bAi$ can be fully expressed with the help of original metric, and four velocity, energy density and pressure of fluid. It is also clear that connection is still compatible with the matrix $\bAi$. However it is clear from the form of the relation between $\bAi^{ab}$ and $g^{ab}$ that is more natural 
to express metric $g^{ab}$ as function of $\bAi^{ab}$ and four velocity
\begin{equation}\label{gabA}
g^{ab}=\frac{1}{X}(\bAi^{ab}-Y u^au^b)
\end{equation}
and hence
\begin{equation}
	\det g=\frac{1}{\det g^{ab}}=X^4\frac{\det \bA}{
	1-Y\tu_a u^a} \ , \quad  \tu_a=\bA_{ab}u^b \ .
\end{equation}
Further, from (\ref{gabA}) we easily obtain inverse matrix 
$g_{ab}$ to be equal to
\begin{equation}
	g_{ab}=X(\bA_{ab}+\frac{Y}{1-Yu^a\bA_{ab}u^b}\tu_a\tu_b) \ , 
\end{equation}
where $Y$ and $X$ are given in 
(\ref{Y}) and (\ref{X}) respectively.
%
Then finally using the fact that $\bA_{ab}=g_{ab}+\frac{1}{M^2_{BI}}(R_{ab}-\kappa \ot_{ab}) $ we obtain equations of motion
\begin{eqnarray}
&&	R_{ab}(\Gamma)-\kappa \ot_{ab}(\bA)=
M_{BI}^2(\bA_{ab}-g_{ab})=\nonumber \\
&&M^2_{BI}(1-X)\bA_{ab}-M^2_{BI}\frac{Y}{1-Y u^a\bA_{ab}u^b}\tu_a\tu_b   \  \nonumber \\
	\end{eqnarray}
that can be  rewritten into the form 
\begin{equation}
	R_{ab}(\bA)-\kappa \mathbf{\ot}_{ab}=0 \ , 
\end{equation}
where 
\begin{eqnarray}
&&	\mathbf{\ot}_{ab}=
[\frac{M^2_{BI}}{\kappa}(1-X)+\frac{1}{2}(\rho-p)X]\bA_{ab}+
\nonumber \\
&&+[(\rho+p)\frac{X^2}{(1-Y u^m\tu_m)^2}+\frac{1}{2}
\frac{(\rho-p)XY}{(1-Y u^m\tu_m)}-\frac{M^2_{BI}}{\kappa}\frac{XY}{1-Y u^m\tu_m}]\tu_a\tu_b
=	\nonumber \\
&&=(\tilde{\rho}+\tilde{p})\tu_a\tu_b+
\frac{1}{2}(\tilde{\rho}-\tilde{p})\bA_{ab} \ , \nonumber \\
\end{eqnarray}
where in the final step we introduced $\tilde{\rho}$ and $\tilde{p}$ that are unequally determined by $X,Y$ and $\rho,p$ and $u^a$.
We see that the equations of motion for $\bA_{ab}$ have the same
form as in the case of standard equations of motion for gravity that interacts with ideal fluid that however has rather  complicated equation of state.  We mean that this is an interesting  result 
which deserves to be elaborated further. For example, it would be nice to analyze cosmological
solutions of this theory for standard matter components in the Universe with well known relation between $\rho$ and $p$. 
\section{Auxiliary Field Representation}\label{fifth}
In order to gain more insight into modified Eddington action (\ref{Padmod}) it is 
instructive to rewrite it  with the help of auxiliary symmetric tensor $\hg_{ab}$ into the form 
\begin{eqnarray}\label{Padmodau}
&&S=M_p^2M_{BI}^2
\int d^4x (\frac{1}{2}\sqrt{-\det \hg}(\hg^{ab}\bA_{ba}-2)-
\lambda \sqrt{-g}) \ , \nonumber \\
\quad &&\bA_{ab}=g_{ab}+\frac{1}{M^2_{BI}}(R_{(ab)}(\Gamma)-\kappa\ot_{ab}) \ . 
\nonumber \\
\end{eqnarray}
To see an equivalence between (\ref{Padmodau}) and (\ref{Padmod})   let us consider equations of motion for $\hg$ that follow from (\ref{Padmodau})
\begin{equation}\label{eqhg}
-\frac{1}{2}\sqrt{-\det\hg}\hg_{ab}(\hg^{cd}\bA_{cd}-2)+
\sqrt{-\det\hg}\bA_{ab}=0 \ . 
\end{equation}
This equation  has solution $\hg_{ab}=\bA_{ab}$. Then inserting this result into 
(\ref{Padmodau}) we recovery the action (\ref{Padmod}) that proves en equivalence of these
two formulations. 

Performing variation of (\ref{Padmodau}) with respect to independent connection again leads to  compatibility condition 
\begin{equation}
	\nabla_c [\sqrt{-\det\hg}\hg^{ab}]=0
\end{equation}
that implies that $\Gamma^c_{ab}$ has the form of Levi-Civita connection for auxiliary metric $\hg_{ab}$ 
\begin{equation}
\Gamma^c_{ab}=\frac{1}{2}\hg^{cr}(\partial_a \hg_{rb}+
\partial_b \hg_{ra}-\partial_r \hg_{ab}) \ .
\end{equation}
Further, variation of the action (\ref{Padmodau}) with respect to $g_{ab}$ leads to the equations of motion for $g_{ab}$  
\begin{equation}
	\frac{1}{2}\sqrt{-\det\hg}\hg^{mn}
	(\frac{1}{2}(\delta_m^a\delta_n^b+\delta_m^b\delta_n^a)
	-\frac{\kappa}{M_{BI}^2}\frac{\delta \ot_{mn}}{\delta g_{ab}})	
	-\frac{\lambda}{2}\sqrt{-g}g^{ab}=0 \ . 
\end{equation}
And finally  variation of action (\ref{Padmodau}) with respect to $\hg_{ab}$ gives equation 
(\ref{eqhg}) that has solution
\begin{equation}
	\hg_{ab}=g_{ab}+\frac{1}{M^2_{BI}}(R_{(ab)}(\Gamma)-\ot_{ab}) \ . 
\end{equation}
In the simplest case of absence of mater we obtain that 
\begin{equation}
	\sqrt{-\det\hg}\hg^{ab}=\lambda \sqrt{-g}g^{ab}
\end{equation}
that implies $\hg^{ab}=\frac{1}{\lambda}g^{ab}$
and hence equations above has the form
\begin{equation}
	R_{ab}=(\lambda-1)M_{BI}^2 g_{ab}
\end{equation}
and we recovery the standard result of Eddington gravity. 

In case of the scalar field we again have
\begin{equation}
	\ot_{ab}=\partial_a\phi^A\partial_b \phi^BK_{AB}-\frac{1}{2}Vg_{ab}
\end{equation}
and we get
\begin{equation}
	\sqrt{-\hg}\hg^{ab}(1+\frac{\kappa}{2M^2_{BI}}V)=
	\lambda\sqrt{-g}g^{ab}
\end{equation}
that has solution
\begin{equation}
	\hg^{ab}=\frac{(1+\frac{\kappa}{2M_{BI}^2}V)}{\lambda}g^{ab} \ , 
	\quad 
	\hg_{ab}=\frac{\lambda}{1+\frac{\kappa}{2M_{BI}^2}V} g_{ab}
	\end{equation}
In term of metric $\hg$ the equations of motion have the form 
\begin{eqnarray}
	R_{ab}(\hg)-\kappa\partial_a\phi\partial_b\phi+
\frac{M^2_{BI}}{\lambda}(1+\frac{\kappa}{2M_{BI}^2})^2\hg_{ab}
-M_{BI}^2\hg_{ab}=0 \nonumber \\	
\end{eqnarray}
that have the same form as equations of motion derived in previous section. We see that it is natural to interpret $\hg_{ab}$ as dynamical metric since their equations of motion are equivalent to equations of motion derived from Einstein-Hilbert action that interacts with scalar field with new potential term. Then  $g_{ab}$ can be considered as an auxiliary tensor that is necessary for definition of stress
energy tensor and also for consistency of the action (\ref{Padmodau}).

It is also clear that we could proceed in the same way with the stress energy tensor for ideal fluid but we will not repeat this analysis here since the result is obvious: The dynamics of modified Eddington gravity interacting with fluid  is more naturally formulated win terms of  the new metric field $\hg_{ab}$ and new stress energy tensor for fluid. It would be interesting to elaborate this idea further for example in case of more general form of matter.

{\bf Acknowledgment:}

This work  is supported by the grant “Dualitites and higher order derivatives” (GA23-06498S) from the Czech Science Foundation (GACR).

\end{document}